\documentstyle[editedvolume,epsf]{crckapb}

\def\scnot#1#2{#1 \times 10^{#2}}

\begin{opening}
\title{SMALL SCALE STRUCTURE AND HIGH REDSHIFT HI}

\author{David H. Weinberg}
\institute{Ohio State University\\
           Dept. of Astronomy\\
	   174 W. 18th Ave.\\
	   Columbus, OH 43210\\
	   USA\\
           }
\author{Lars Hernquist}
\institute{U.C. Santa Cruz\\
	   Dept. of Astronomy\\
	   Santa Cruz, CA 95064\\
	   USA\\
	   }
\author{Neal S. Katz}
\institute{University of Washington\\
	   Dept. of Astronomy\\
	   Seattle, WA 98195\\
	   USA\\
	   }
\author{Jordi Miralda-Escud\'e}
\institute{Institute for Advanced Study\\
	   Olden Lane\\
	   Princeton, NJ 08540\\
	   USA\\
	   }

\end{opening}

\runningtitle{SMALL SCALE STRUCTURE AND HIGH REDSHIFT HI}

\begin{document}

\noindent
To appear in {\it Cold Gas at High Redshift}, eds. M. Bremer, H. Rottgering,
C. Carilli, and P. van de Werf, Kluwer, Dordrecht (1996)

\newcommand{\kms}{km \hskip -2pt s$^{-1}$}
\newcommand{\hubunits}{km \hskip -2pt s$^{-1}$ \hskip -2pt Mpc$^{-1}$}
\newcommand{\lya}{Ly$\alpha$}
\newcommand\junits{{\rm erg\,s}^{-1}\,{\rm cm}^{-2}\,{\rm sr}^{-1}\,
		   {\rm Hz}^{-1}}
\newcommand\cdunits{{\rm cm}^{-2}}
\newcommand{\K}{{\rm K}}

\newbox\grsign \setbox\grsign=\hbox{$>$} \newdimen\grdimen \grdimen=\ht\grsign
\newbox\simlessbox \newbox\simgreatbox
\setbox\simgreatbox=\hbox{\raise.5ex\hbox{$>$}\llap
     {\lower.5ex\hbox{$\sim$}}}\ht1=\grdimen\dp1=0pt
\setbox\simlessbox=\hbox{\raise.5ex\hbox{$<$}\llap
     {\lower.5ex\hbox{$\sim$}}}\ht2=\grdimen\dp2=0pt
\newcommand{\simgt}{\mathrel{\copy\simgreatbox}}
\newcommand{\simlt}{\mathrel{\copy\simlessbox}}

\section{Introduction}

Galaxy redshift surveys reveal the presence of large scale structure
in the local universe, a network of sheets and filaments interlaced
with voids and tunnels.  Zel'dovich \shortcite{Zel'dovich70} showed
that gravitational instability in an expanding universe can create
such structures from generic random initial conditions.
Zeldovich's analysis was originally used to describe the first
collapse in ``top--down'' scenarios like the hot dark matter
model, which have a cutoff in the primordial
fluctuation power spectrum at small scales.
The theories of structure formation that are most popular today
have no intrinsic cutoff in the power spectrum.
Structures in such a theory grow by hierarchical clustering ---
low mass perturbations collapse early, then merge into progressively
larger objects.

The Zel'dovich analysis does not apply directly to hierarchical
clustering models, but numerical and analytic studies show that
they tend to develop the same types of structure
(e.g., \citeauthor{Shandarin89}, \citeyear{Shandarin89};
\citeauthor{Weinberg90}, \citeyear{Weinberg90};
\citeauthor{Melott93}, \citeyear{Melott93}).
The smooth ``pancakes'' of the top--down theory are replaced
by ``second generation pancakes'' that are themselves made up
of smaller clumps. The characteristic scale of voids, sheets,
and filaments grows with time, as larger scales reach the
non-linear regime.  In a hierarchical scenario, one naturally
expects the high redshift universe to contain ``small scale structure''
that is qualitatively similar to today's large scale structure,
but reduced in size by a factor that depends on the specifics
of the cosmological model.

Observations of absorption and emission by neutral hydrogen can
trace this small scale structure over a wide range of redshifts.
Such observations probe the evolution of the intergalactic medium
and the condensation of gas into galaxies, filling in the gap
between cosmic microwave background anisotropies and maps of
present day structure.
On the theoretical side, an important recent development is
the use of hydrodynamic simulations to work out the predictions
of {\it a priori} cosmological models for observable high redshift
structure.  This talk is based primarily on the results of a
numerical simulation of the cold dark matter (CDM) model using
TreeSPH, a combined N-body/hydrodynamics code.
The simulation methods and some applications to galaxy formation
are discussed in Katz {\it et al.} \shortcite{Katz95a},
and some early results on Ly$\alpha$ absorbers are described
in Katz {\it et al.} (\citeyear{Katz95b}, hereafter KWHM)
and Hernquist {\it et al.} (\citeyear{Hernquist95}, hereafter HKWM).

\section{Ly$\alpha$ Absorption in the CDM Model}

% Figure 1: box= 135 415 425 720

% Figure 2: box= 60 485 320 740

% Figure 3: box= 85 235 555 735

% Figure 4: box= 75 250 510 685

% Figure 5: box = 45 180 565 650

% Figure 6: box = 90 410 460 720

\begin{figure}
\epsfxsize=4.8in
\centerline{
\epsfbox[135 415 425 720]{figSim.ps}
}
\caption{\label{figSim}
The distribution of gas particles in a hydrodynamic simulation of the
CDM model, at $z=2$.  The simulation volume is a cube 22.222 comoving
Mpc on a side (for $h=0.5$), making the physical size at
this redshift 7.4 Mpc.
}
\end{figure}

Figure~\ref{figSim} shows the distribution of gas (SPH) particles at $z=2$
in a simulation of the ``standard'' CDM model, with
parameters $\Omega=1$, $\Omega_b=0.05$, and
$h\equiv H_0/100\,$\hubunits$=0.5$.
The simulation volume is a periodic cube of comoving size
22.222 Mpc, so its physical size at $z=2$ is 7.4 Mpc, with a
corresponding Hubble flow of 1925 \kms.  There are $64^3$
SPH particles to represent the baryon component and $64^3$
collisionless particles (not shown) to represent the
cold dark matter component; individual particle masses are
$\scnot{1.5}{8}M_\odot$ and $\scnot{2.8}{9}M_\odot$, respectively.
The simulation incorporates radiative cooling for a gas of
primordial composition (76\% hydrogen, 24\% helium) in ionization
equilibrium with an ultraviolet (UV) radiation background of intensity
$J(\nu)=10^{-22}F(z)(\nu_L/\nu)\,\junits$, where
$\nu_L$ is the Lyman limit frequency and
$F(z)=0$ for $z>6$, $4/(1+z)$ for $6>z>3$, and 1 for $3>z>2$.

We normalize the CDM power spectrum so that, if it were linearly
extrapolated to $z=0$, the rms mass fluctuation in spheres of
radius 16 Mpc would be $\sigma_{8h^{-1}{\rm Mpc}}=0.7$.  This normalization
is roughly that required to match the observed abundance of
massive galaxy clusters \cite{White93}.
However, with this normalization and the other parameters
we have adopted, the CDM model predicts large scale microwave
background fluctuations nearly a factor of two lower than
those observed by COBE \cite{Bunn95}.
An $\Omega=1$ model dominated by cold dark matter must
involve some additional complication (e.g. a ``tilted'' or
``broken'' primeval power spectrum, a lower Hubble constant,
an admixture of massive neutrinos)  in order to account for
COBE fluctuations and galaxy clusters simultaneously.
The version of CDM that we have simulated might be a useful
approximation to such a model on the scales considered here.
We plan to examine alternative scenarios --- in particular
low--$\Omega$ CDM models --- in the near future.

The spatial structure in Figure~\ref{figSim} has the
filamentary character seen in typical simulations (and observations)
of large scale structure.  However, the size of the structures is
relatively small --- the largest low density regions, for instance,
have a diameter of $5-10$ comoving Mpc.
This scale would be somewhat larger if the simulation box
were itself large enough to accommodate longer wavelength modes,
but primarily the reduced scale of structures reflects the
lower amplitude of fluctuations at $z=2$ relative to $z=0$.
Only at later times do larger scale fluctuations reach the
amplitude required to produce non-linear gravitational collapse.
At the level of detail discernible in Figure~\ref{figSim}, the
dark matter distribution would look very similar to the depicted
gas distribution.

\begin{figure}
\epsfxsize=3.5in
\centerline{
\epsfbox[60 485 320 740]{figRhoT.ps}
}
\caption{\label{figRhoT}
The distribution of gas in the density--temperature plane at $z=2$.
Each point represents a single SPH particle; temperatures are in
degrees Kelvin and densities are scaled to the mean baryon density.
Histograms show the 1-d marginal distributions, i.e. the fraction of
particles in each decade of density and of temperature.
}
\end{figure}

\begin{figure}
\epsfxsize=4.8in
\centerline{
\epsfbox[85 235 555 735]{figGas.ps}
}
\caption{\label{figGas}
The spatial distribution of gas in different regimes of
density and temperature, as indicated above each panel.
}
\end{figure}

Figure~\ref{figRhoT} shows the distribution of gas in the
density--temperature plane.  Each point represents a single SPH particle,
and histograms at the edges of the Figure show marginal distributions.
This representation reveals four main components.  One is low density,
low temperature gas, which occupies a well defined locus along
which adiabatic cooling balances heating by photoionization.
A second is overdense, shock heated gas; at this redshift, 10\%
of the gas has $T>10^5\,\K$ and 5\% has $T>10^6\,\K$.
The third component consists of very overdense gas that
has radiatively cooled to the equilibrium temperature, $T \approx 10^4\,\K$,
where heat input and radiative cooling balance.
The fourth component is warm gas at moderate overdensity.
While this category is to some extent a ``catch--all'' for gas
that does not fit into one of the other, more distinct components,
it accounts for an appreciable fraction of the baryonic mass.

\begin{figure}
\epsfxsize=4.8in
\centerline{
\epsfbox[75 250 510 685]{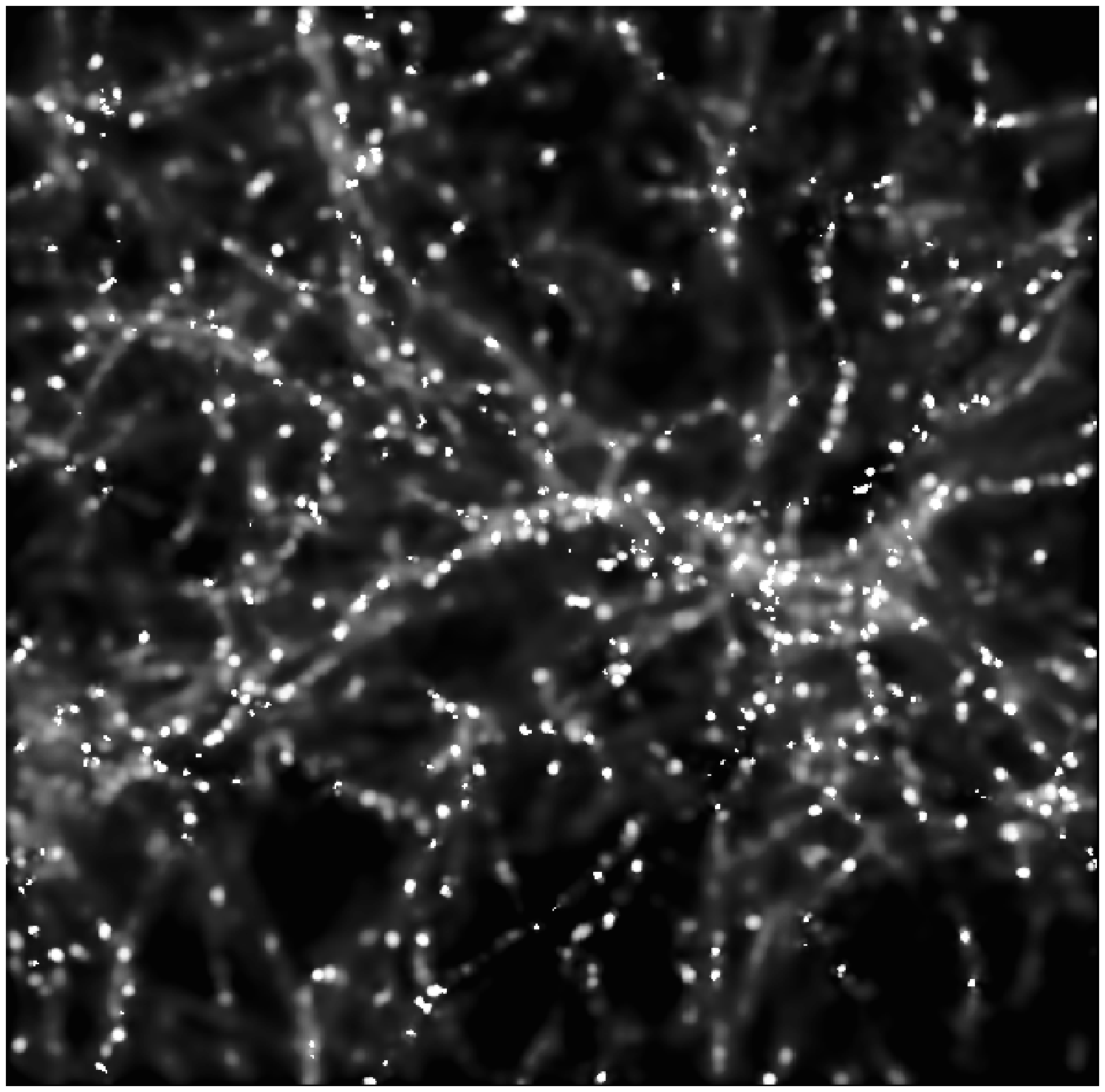}
}
\caption{\label{figMap}
A map of the projected neutral hydrogen column density in the simulation
at $z = 2$.  At this redshift in the $\Omega=1$ cosmology,
the depth of the 7.4 Mpc simulation box would be $c\Delta z = 1924\,$\kms
and the angular size would be 15.1 arc-minutes.
In this representation, the saturated (white) regions have
$N_{\rm HI} \simgt 10^{16.5}\ {\rm cm}^{-2}$, and the faintest
visible structures have $N_{\rm HI} \sim 10^{14.5}\ {\rm cm}^{-2}$.
Even regions that are black in this map can give rise to absorption
at column densities typical of the \lya\ forest.
}
\end{figure}

Figure~\ref{figGas} shows the spatial distribution of the gas in
different regimes of density and temperature.  Gas with
$T<30,000\,\K$ and overdensity $\rho/\bar\rho<1000$ (upper left panel)
mostly occupies the low density regions, though hints of the filaments
in Figure 1 can be seen here as well.
The filaments stand out dramatically in the warm gas component,
with $30,000\,\K < T < 10^6\,\K$ (upper right panel).
This temperature cut selects gas that has been heated by
adiabatic compression and mild shocks as it falls into
moderate overdensity structures.
The hottest gas ($T>10^6\,\K$, lower left panel) is confined
to fully virialized dark matter potential wells, and its
spatial distribution is more clumpy.
The gas with $T<30,000\,\K$ and $\rho/\bar\rho>1000$ (lower right panel)
occupies radiatively cooled knots inside these hot gas halos.
The larger halos may contain several such knots.
The most massive knots contain several hundred particles
(merged into a single extended dot at the resolution of Figure~\ref{figGas}),
while the least massive, which are generally the ones that have
started to cool and condense most recently, contain only a handful
of cold gas particles.
The gravitational softening of the simulation, 7 kpc at $z=2$,
prevents us from resolving the detailed internal structure of
these knots, but the physical conditions imply that they are
likely to fragment and form stars.  It is plausible to identify
these knots as young --- in some cases just forming --- galaxies.

Knowing the density and temperature of each gas particle and the
intensity of the model UV background, we can compute the corresponding
neutral hydrogen fractions assuming ionization equilibrium.
Figure~\ref{figMap} shows a 2-d map of the neutral hydrogen column
density projected through the simulation cube.  There is a close
correspondence between the prominent structures in this map
and the gas distributions in the right hand panels of Figure~\ref{figGas}.
The cold gas knots nearly always produce absorption at a column
density of $N_{HI}=10^{17}\,\cdunits$ or greater, features that
appear white in the grey scale representation of Figure~\ref{figMap}
(these regions are corrected for self-shielding using a procedure
described in KWHM).
The warm gas filaments produce the extended, lacy structures at
lower column density.  It is important to note that the faintest visible
structures in Figure~\ref{figMap} have $N_{HI} \approx 10^{14.5}\,\cdunits$,
so much of the absorption at column densities typical of the
\lya\ forest arises in regions that are black in this Figure.

\begin{figure}
\epsfxsize=4.8in
\centerline{
\epsfbox[45 445 565 650]{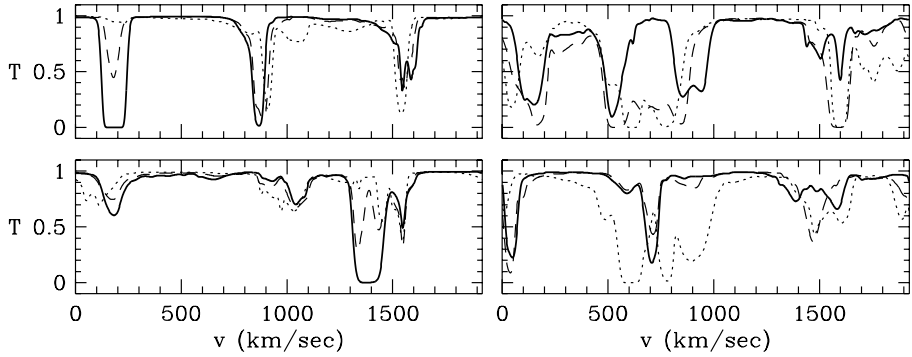}
}
\caption{\label{figSpec}
Examples of artificial spectra at $z=2$.  Solid lines show
transmission against velocity along four random lines of sight.  At
this redshift, the physical size of the periodic simulation box is
7.41 Mpc, corresponding to a Hubble flow of 1924.5 km/s.  Dashed and
dotted lines show spectra along lines of sight displaced arbitrarily
from that of the primary spectrum by physical separations of
100 kpc and 300 kpc, respectively.
The translation from velocity $v$ to wavelength $\lambda$ is
$\lambda=1216 \times (1+z) \times (1+v/c)$ \AA, where $z=2$.
}
\end{figure}

\begin{figure}
\epsfxsize=3.5in
\centerline{
\epsfbox[90 410 460 720]{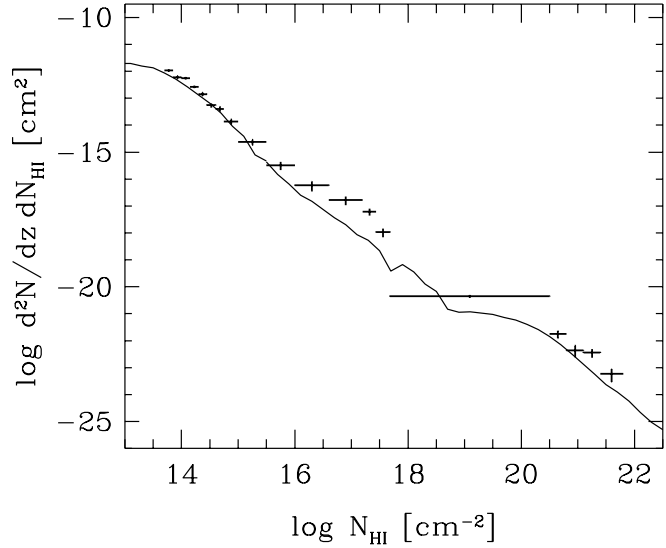}
}
\caption{\label{figColumn}
Distribution of neutral hydrogen column densities.
The solid line shows the simulation results at $z=2$.
Points with error bars are taken from
the observational compilation by Petitjean {\it et al.} (1993).
}
\end{figure}

Figure~\ref{figSpec} shows artificial QSO absorption spectra
along four randomly chosen lines of sight through the simulation
box at $z=2$.  In each panel, the solid line shows the
transmission $T=e^{-\tau}$, where $\tau$ is the \lya\ optical depth.
The dashed line shows a
spectrum along a line of sight 100 kpc away from this primary spectrum.
The dotted line shows a
spectrum at 300 kpc separation (200 kpc from the dashed spectrum).
Many absorption features appear in both of the first two spectra, and
there are significant matches even for lines of sight separated by 300
kpc, though these are
often accompanied by substantial changes in the features'
depth or shape.
Figure~\ref{figMap} shows that the low column density absorbing
structures are extended and coherent, so the correlation of
features along neighboring lines of sight is not surprising.
Qualitatively, it appears that our results can
account for the large coherence scale found in absorption studies of
QSO pairs (e.g. \citeauthor{Bechtold94} \citeyear{Bechtold94};
\citeauthor{Dinshaw94} \citeyear{Dinshaw94}, \citeyear{Dinshaw95}),
though quantitative tests (e.g. \citeauthor{Charlton95} \citeyear{Charlton95})
are needed to assess the agreement or lack of agreement with recent
observations.

Figure~\ref{figColumn} displays the distribution of neutral hydrogen
column densities, $d^2N/dN_{HI}\,dz$, at $z=2$.
The procedure adopted to identify lines at column densities
$N_{HI}<10^{15.5}\,\cdunits$ is described in HKWM.
Above this column density, we determine the distribution
by measuring the fractional area above each column density in the
projected HI map of Figure~\ref{figMap}.
High column density lines are rare enough
that the absorption along any line of sight that has
$N_{\rm HI}>10^{15.5}\,\cdunits$ through our 7.4 Mpc box
is always dominated by a single absorber.
Observational data and error bars in Figure~\ref{figColumn}
are taken from Table 2 of
\citeauthor{Petitjean93} \shortcite{Petitjean93}.
There is a significant (factor of ten) discrepancy with the
\citeauthor{Petitjean93} data for column densities near
$10^{17}\cdunits$, which could reflect
either a failure of standard CDM or the presence in the real universe
of an additional population of Lyman-limit systems that are not
resolved by the simulation.  Nonetheless, given that CDM is an
{\it a priori} theoretical model that was not ``designed'' or
adjusted to fit these observations,
the overall level of agreement across eight orders of magnitude
in neutral hydrogen column density is rather remarkable.
Other analyses of absorption in this simulation are presented
by HKWM and KWHM.

\citeauthor{Cen94} \shortcite{Cen94}, who were the first to use
these sorts of simulations to model the \lya\ forest,
report similar agreement between observations and a low-$\Omega$
CDM model with a cosmological constant.
\citeauthor{Zhang95} \shortcite{Zhang95} find similar agreement
for an $\Omega=1$ CDM model with a higher normalization
($\sigma_{8h^{-1}{\rm Mpc}}=1$) than used here.
The qualitative success of three different models (and numerical
methods) suggests that the \lya\ forest arises naturally,
and at least somewhat generically, in a hierarchical theory
of structure formation with a photoionizing background.
The comparison between simulations and the extraordinary data
emerging from high resolution QSO spectra can clearly be carried
out much more carefully than has been done so far.
There is every reason to hope that detailed comparisons will
reveal discrepancies that restrict the pool of acceptable
theoretical models.  For now, the agreement between simulated
and observed line populations suggests that the simulation
described here is worth taking seriously as a realistic
general picture for the origin of \lya\ absorbers.

In this simulation, the structures that produce
low column density absorption ($N_{\rm HI} \sim 10^{13}-10^{15}\cdunits$)
are physically diverse: they include
filaments of warm gas, caustics in frequency
space produced by converging velocity flows \cite{McGill90},
high density halos of hot, collisionally ionized gas,
layers of cool gas sandwiched between shocks \cite{Cen94},
and modest local undulations in undistinguished regions of the
intergalactic medium.  Temperatures of the absorbing gas range from
below $10^4 K$ to above $10^6 K$.
The ``typical'' low column density
absorbers --- to the extent that we can identify such
a class --- are flattened structures of rather low overdensity
($\rho/\bar\rho \sim 1-10$), and their line widths are often
set by peculiar motions or Hubble flow rather than thermal broadening.
Because of their low overdensities,
most of the absorbers are far from dynamical or thermal equilibrium,
and many are still expanding with residual Hubble flow.
The simulation reveals a smoothly fluctuating intergalactic medium,
with no sharp distinction between ``background'' and ``\lya\ clouds''.
Indeed, in this picture one might say that the \lya\ forest {\it is}
the absorption by diffuse intergalactic hydrogen known as
the Gunn-Peterson \shortcite{Gunn65} effect.
There is also absorption by low density gas outside of the ``lines,''
but it is relatively weak.  At higher redshifts, increasing neutral
fractions make the absorbing gas more opaque, and the distinction
between lines and background becomes even harder to draw.

A comparison between Figures~\ref{figGas} and \ref{figMap} shows
a clear association between high column density absorbers and the
knots of radiatively cooled gas that represent forming galaxies.
Damped \lya\ absorption ($N_{\rm HI} \geq 10^{20.2}\,\cdunits$)
occurs along lines of sight that pass through the denser, more massive
protogalaxies.  The column density correlates inversely with
the projected distance from the protogalaxy center, and
the maximum projected separation that yields damped absorption
is about 20 kpc (see KWHM).
Lyman limit absorption
($10^{17}\,\cdunits \leq N_{\rm HI} \leq 10^{20.2}\,\cdunits$)
occurs on lines of sight that pass either through the outer
parts (20--100 kpc projected separation) of the more massive
protogalaxies or near the centers of younger, lower density systems.
By definition the stronger Lyman limit systems are self-shielded
against the ionizing background, so they are mainly neutral,
in constrast to the \lya\ forest systems, which usually have
neutral fractions of $10^{-6}-10^{-4}$.

\section{Prospects for 21cm Observations}

Figure~\ref{figMap} can be regarded as a (highly) idealized 21cm observation
of a CDM universe at $z=2$, covering a region 15.1 arc-minutes on a side
and 1924 \kms\ in velocity.
The contributions of Ingram and Braun to these proceedings show
examples of what the Square-Kilometer Array (SKA) might see if
it observed a universe like this, with realistic assumptions about
resolution and noise.  Here we will stick to more
generic comments, inspired by the simulation but not directly
tied to it.
Some of the issues that affect the observability of high redshift HI
in a variety of cosmological models are discussed by
Scott and Rees \shortcite{Scott90} and references therein.

In a typical hierarchical model, the ``re''combination of hydrogen
at $z=1100$ is followed by an era of linear growth, with the
density contrast of fluctuations $|\delta|\ll 1$ on all cosmologically
relevant scales.  To the extent that pressure and Compton drag against
the microwave background can be ignored, density contrasts grow in
proportion to the expansion factor, $\delta \propto a \propto t^{2/3}$.
Once the strongest fluctuations (which occur on small scales) reach
$\delta \sim 1$, they collapse, eventually giving rise to luminous
objects that reionize the remaining diffuse gas.  It is not at all clear
what objects actually cause reionization (globular clusters? quasars?
supermassive stars? dwarf galaxies?), but the visibility of high
redshift quasars at wavelengths shorter than \lya\ implies that
the reionization occurred before $z \approx 5$.
It could plausibly have happened much earlier
(\citeauthor{Tegmark94}, \citeyear{Tegmark94};
\citeauthor{Fukugita94}, \citeyear{Fukugita94}),
though the detection of microwave background anisotropies at degree
scales suggests that it did not occur at $z$ much above 50
\cite{Scott95}.

Reionization heats the diffuse medium to $T \sim 10^4\,\K$, introducing
a Jeans mass --- thermal pressure prevents baryons from collapsing into
objects with circular velocities below about 35 \kms
(\citeauthor{Quinn95}, \citeyear{Quinn95};
\citeauthor{Thoul95}, \citeyear{Thoul95}).
Depending on the relative timing of reionization and fluctuation
growth, the universe may enter a quiet phase during which little
collapse occurs.  Eventually, however, fluctuations larger than
the Jeans mass reach the nonlinear regime, and the condensation of
gas into protogalaxies and consequent star formation can begin in
earnest.  As time goes on and larger scale fluctuations become nonlinear,
galaxies pull themselves into groups, clusters, and superclusters.

Reionization may have occurred at a redshift beyond the reach of
current radio telescopes.  However, it could have occurred at a
redshift only slightly greater than 5, since at $z=5$ the density of quasars
is plummeting
and the opacity of the diffuse medium is rising.  In this case,
there might be interesting observables at $5<z<8$.
Structure should be present on Mpc scales, and much of the gas
should be cool enough to have a high neutral fraction.
More work, including analysis of simulations, is needed to
investigate whether the predicted structure is bright enough to
detect with existing or plausible future instruments.
If reionization takes place by the growth of ``Stromgren spheres''
around quasars or other rare objects, then there might be strong
structure in the neutral hydrogen even where the underlying gas
distribution is fairly uniform.

The prospects for HI emission searches above $z=5$ are intriguing
but highly uncertain.  Below this redshift, observations give more
guidance about what to expect.  In particular, we know that the
universe is reionized, and the statistics of \lya\ absorption give
an estimate of the neutral hydrogen density parameter $\Omega_{HI}$.

In the simulation of \S 2, most of the gas at $z \sim 2-4$ is either
in the low density, photoionized medium (roughly speaking, the
\lya\ forest), or in the high density, hot, collisionally ionized medium.
However, nearly all of the {\it neutral} gas is in high density,
radiatively cooled objects --- Lyman limit and damped \lya\ absorbers.
This theoretical result is consistent with observational inferences,
which show that the value of $\Omega_{HI}$ is dominated by contributions
from the highest column density systems.

Single objects containing $10^{14}M_\odot$ of HI do not appear in
the simulation, and they seem unlikely on generic grounds.
The collapse of an object with $10^{14}M_\odot$ of baryons would
shock--heat and collisionally ionize the gas, and before collapse
the gas would be low density and photoionized.  One could in
principle have an object in which $10^{14}M_\odot$ of baryons
collapsed {\it and} cooled --- a sort of super damped \lya\ system ---
but if such objects were common we would expect to see many
galaxies with $10^{14}M_\odot$ of stars today.

The simulated observations of Ingram and Braun in these proceedings show that
it is difficult to detect the faint caustics seen in Figure~\ref{figMap}
even with an instrument as ambitious as the SKA.  What the SKA
can detect rather easily are the damped \lya\ systems, which are more
numerous than present day, $L^*$ galaxies by a factor of several.
A single pointing covers a small angular field, but it is sensitive
to a gigantic range in redshift.  A program of multiple long exposures
could thus provide a ``galaxy'' (damped \lya) redshift survey over
a large cosmological volume at high redshift.  This would be an
extraordinary feat, and observing the history of galaxy clustering
would doubtless teach us a great deal about the underlying physics
of structure formation.

At lower sensitivity, the first objects to be detected in 21cm
emission at high redshift are likely to be clusters of damped \lya\ systems,
the high redshift analogs of today's rich galaxy clusters.  The
statistics of these can in principle be calculated from simulations,
but very large simulation volumes are needed to predict the abundance
of the rarest, most massive systems.  We will therefore attempt a
more phenomenological calculation, based on scaling the abundance
of present galaxy clusters back to high $z$.

\citeauthor{Bahcall93} \shortcite{Bahcall93} find that the abundance
of observed rich clusters with total mass $M$ or greater
can be described by the cumulative mass function
\newcommand{\nos}{n_0^*}
\newcommand{\mos}{M_0^*}
\newcommand{\nzs}{n_z^*}
\newcommand{\mzs}{M_z^*}
$$
n(>M) = \nos(M/\mos)^{-1} e^{-M/\mos},
$$
with $\mos=\scnot{1.8}{14}h^{-1} M_\odot$ and
$\nos=\scnot{4}{-5} h^3$ Mpc$^{-3}$.
Suppose that this mass function has emerged from a hierarchical
model of structure formation in which the power spectrum on the
scales of interest can be approximated as $P(k) \propto k^n$.
The characteristic mass $\mos$ corresponds to the scale
on which the rms linear theory mass fluctuations have some
value $\sigma^* \sim 1/2$, so that only the rare ($\sim 2-3\sigma$)
fluctuations on this scale have collapsed to form virialized clusters.
At high redshift we might expect a mass function of similar form,
but the characteristic mass $\mzs$ will correspond to the (smaller)
scale on which the rms fluctuation is $\sigma^*$ at this redshift.
For a $k^n$ power spectrum,
$$
\sigma(M,z) = \sigma^*(M/\mos)^{-(3+n)/6} D(z),
$$
where a useful approximation to the growth factor $D(z)$ is
\cite{Shandarin83}
$$
D(z) \approx \left(1 + {2.5\Omega_0 z \over 1+1.5\Omega_0}\right)^{-1}.
$$
The condition $\sigma(\mzs,z)=\sigma^*$ implies that the
characteristic mass at redshift $z$ is $\mzs=\mos D^{6/(3+n)}.$
The abundance $\nzs$, in {\it comoving} units, can be determined
from the condition $\nzs\mzs=\nos\mos \Longrightarrow
\nzs=\nos D^{-6/(3+n)}$, since objects of mass $M>M^*$ contain
a constant fraction of the total mass of the universe in this
sort of scale free model.

For our purposes, we are interested in a cluster's HI mass rather
than its total mass.  To go from one to the other, we can make the
adventurous assumption that the ratio of $M_{HI}$ to $M_{tot}$ is
the same as the universal ratio $\Omega_{HI}(z)/\Omega(z)$.
This assumption is almost certainly incorrect today because the
galaxies in clusters tend to be gas poor, early types.  It may
be more plausible at high redshift, when galaxies are primarily
gaseous; indeed, the error may be in the opposite sense
(underestimating $M_{HI}$ instead of overestimating) if galaxies
at high $z$ formed preferentially in the densest regions.
Putting all of this together, we arrive at an expression for
the cumulative, comoving number density of clusters with HI mass
greater than $M_{HI}$:
$$
n(>M_{HI}) = \nos F
\left({M_{HI} F \over R M_0^*} \right)^{-1}
{\rm exp}\left({-M_{HI} F \over R\mos}\right)~,
$$
$$
\mos=\scnot{1.8}{14}h^{-1} M_\odot, \quad
\nos=\scnot{4}{-5} h^3 {\rm \ comoving\ Mpc}^{-3},
$$
$$
F \equiv D^{-6/(3+n)}, \quad
R \equiv {\Omega_{HI}(z) \over \Omega(z)} =
{\Omega_{HI}(z) \over \Omega_0} {(1+\Omega_0 z) \over (1+z)}~.
$$

There are many ways that this calculation could depart from reality,
but it illustrates how the power spectrum, the cosmological
model, and the history of the neutral gas density might interact in
determining the abundance of observable high redshift objects.
Unfortunately, the numbers that it yields are not particularly
encouraging because even today the HI mass of an $M^*$ object
is only several$\,\times\, 10^{11}h^{-1}M_\odot$, and at
higher redshift the tendency of galaxies to be more gas rich is
countered by the lower masses of the largest collapsed clusters.
As a specific example, if we adopt $z=3$, $n=-1$, $\Omega_0=0.3$,
and $\Omega_{HI}(z)=0.004$, then $F=16.61$, $R=\scnot{6.3}{-3}$,
the HI mass of an $M^*$ object is
$R\mos F^{-1} = \scnot{6.9}{10} h^{-1} M_\odot$,
and the abundance of such objects is
$\nos F e^{-1} = \scnot{2.4}{-4}\, h^3$ comoving Mpc$^{-3}$.
For an HI mass of $\scnot{3}{11} h^{-1} M_\odot$, 4.37 times higher,
the abundance is down by a factor of $e^{-4.37}/4.37$ to
$\scnot{6.9}{-7} h^3$ comoving Mpc$^{-3}$.

Of course, the fact that this argument leads to a low abundance of
massive HI concentrations means that detection of such a concentration
would be all the more interesting.  The most dubious elements of the
argument (if one is looking for order of magnitude gains, not
factors of 2) are probably the assumption that the HI to total mass
ratio $R$ is universal and the assumption that the objects easiest to
detect are indeed collapsed clusters as opposed to, e.g., lower overdensity
structures that are just detaching from the Hubble flow.

\section{Conclusions}

According to conventional theories of cosmic structure fomration,
the large scale structure that we observe today should be
mirrored in a scaled down form at high redshift.
Absorption and emission measurements of high redshift HI can
trace out the elements of this small scale structure.
The agreement between hydrodynamic simulations and observed
quasar spectra suggests that the \lya\ forest is produced
largely by the moderately overdense ($\rho/\bar\rho \sim 1-10$)
components of this structure, especially the collapsing filaments
and sheets of warm, photoionized gas.
Lyman limit and damped \lya\ absorption probably arises in the
radiatively cooled gas of forming galaxies.
Detecting 21cm emission from high redshifts is an ambitious goal,
but simulated observations and inferences from absorption suggest
that the radio arrays of the future could map the youthful universe
in the way that today's galaxy
redshift surveys have mapped the local large scale structure.
The opportunity to watch galaxies, clusters, voids, and superclusters
grow through time should take us a long way towards understanding
their origin in the physics of the big bang.

\medskip
DW acknowledges research and travel support from NASA grant NAG5-2882.

\end{document}